# Ultimate performance of Quantum Well Infrared Photodetectors in the tunneling regime

E. Lhuillier,[1,2] I. Ribet-Mohamed,[1] M. Tauvy,[1] A. Nedelcu.,[3] V. Berger,[2] E. Rosencher[1]

[1]*ONERA, Chemin de la Hunière, 91761 Palaiseau cedex, France.*

[2]*Matériaux et Phénomènes Quantiques, Université Paris 7, Bat. Condorcet, Case 7021, 75205 Paris cedex 13, France.*

[3]*Alcatel-Thales III-V Lab, Campus de Polytechnique, 1 Avenue A. Fresnel, 91761 Palaiseau cedex, France.*

**Abstract**

Thanks to their wavelength diversity and to their excellent uniformity, Quantum Well Infrared Photodetectors (QWIP) emerge as potential candidates for astronomical or defense applications in the very long wavelength infrared (VLWIR) spectral domain. However, these applications deal with very low backgrounds and are very stringent on dark current requirements. In this paper, we present the full electro-optical characterization of a 15µm QWIP, with emphasis on the dark current measurements. Data exhibit striking features, such as a plateau regime in the I(V) curves at low temperature (4 to 25 K). We show that present theories fail to describe this phenomenon and establish the need for a fully microscopic approach.

## 1. Introduction

Quantum well infrared photodetectors (QWIP) provide nowadays a well established technology to realize infrared focal plane arrays in the 8-12µm spectral range. Indeed, these detectors rely on the use of GaAs materials, and thus benefit from the maturity of the III-V molecular beam epitaxy (MBE) growth and processing techniques. Focal plane arrays as large as 1024x1024 pixels are now available [1, 2], with an excellent uniformity and a very low defect density. In addition, the excellent GaAs material quality results in the absence of 1/f noise [3], making dynamic two point correction unnecessary. Moreover, the peak wavelength of QWIP can easily be tuned thanks to band gap engineering, that is by changing the wells width and the aluminium concentration in the barriers. This flexibility allows QWIP to address the very long wavelength infrared (VLWIR) domain, i. e. wavelengths higher than 12µm [3, 4].

Contrary to the well known infrared atmospheric windows (3-5µm and 8-12µm) which are generally used for ground-based applications, the VLWIR domain can be very useful for space-based applications, either in the astronomy [3] or in the defense domain. These applications call for very specific needs in terms of infrared focal plane arrays. For example, as far as space surveillance and space situational awareness are concerned [3,5], most of the information on target relies on radiometry (and not on imagery). As a consequence, excellent uniformity and linearity are required; the needs in terms of pixel operability are also important, since degraded or dead pixels could miss a target. With their excellent uniformity and high pixels operability [6], QWIP could emerge as potential candidates for these applications. But space surveillance and space situational awareness often require the observation and the tracking of very faint objects against very dark backgrounds. As a consequence, the internal detector noise must be extremely low, in order to remain lower than the background equivalent noise. This places very stringent requirements on the dark current.

During the past decade, much work has been devoted to reducing the dark current of 8-9µm QWIP, and thus to increase their operating temperature up to 75K [7]. But the issue of VLWIR detectors is totally different: indeed, in the applications considered here, a very low operating temperature is generally not an operational problem, so that the detectors are no longer operated in the thermionic regime, but in the tunnel regime. Paradoxically, apart from the early work of Levine *et al* in strongly tunnel-coupled QWIPs [8], this regime, which determines the ultimate performance of QWIP, has never been studied in detail, mainly because of both theoretical and experimental difficulties.

In this paper, we report the full electro-optical characterisation of a 15µm QWIP realised by Alcatel-Thales III-V Lab. The components are described in section 2, and the experimental setup in section 3. The dark current measurements are presented in section 4, while section 5 gives complementary measurements, namely spectral response, responsivity and noise. Section 6 compares the dark current measurements with present theories. Section 7 studies the possibility to attribute the current plateau to electric field domain formation.

## 2. Samples details

The QWIP detectors studied here were designed to have a peak wavelength at 14.5µm. They rely on a forty periods structure, with 7.3nm ($L_w$) GaAs wells and 35nm ($L_b$) $Al_{15.2}Ga_{84.8}As$ barriers. The wells are n type Si doped in their central third with a concentration of $3.10^{11}cm^{-2}$. The structure is sandwiched between two n type contacts with a Si concentration of $10^{18}cm^{-3}$. QWIP detectors were processed into mesa. The sample is a matrix of 384*288 pixels of 25µm pitch (pixel size 23.5µm) in which some pixels are addressed individually. This specific design guarantees that the results obtained on these pixels will be representative of the performance of a focal plane array. The electromagnetic coupling is realized by a grating etched on top of the pixel, with a period of 4.4µm, i.e. very close to the peak wavelength in the material. The potential wells are thus 127meV deep, the ground state (energy: $E_1$) is located 40meV above the bottom of the GaAs conduction band. The Fermi level (energy: $E_{fw}$) is 10.6meV above the ground state. The excited state (energy: $E_2$) is quasi resonant with the top of the barrier (energy $V_b$) since it is only 2meV below the continuum.

## 3. Experimental Setup

All the measurements were done with the QWIP detectors on the cold finger of a Janis continuous flow helium cryostat. A Lakeshore 330 controller was used to reach the desired temperature. To perform measurements in the tunnel regime, it is necessary to cool down the detectors at any desired temperature between 4K and 60K, with i) an excellent stability in time, ii) a very good reproducibility, iii) an accurate measurement of the detectors temperature and iv) a very low noise level. For this purpose, a dedicated cold shield was designed, and three DT 470 temperature probes were implemented: on the cold finger, on the sample holder and on the sample itself, respectively. To measure very low currents, we used a 6430 Keithley sub-femto amperemeter, which allowed us to measure currents as low as a few tens of fA with this experimental setup. Particular care was also devoted to limit parasitic noise by using shielded wires and by getting rid of the surrounding 50Hz noise.

This experimental setup allowed us to perform detailed dark current measurements which will be presented in section 4. It was also used to perform complementary measurements, namely: spectral response, responsivity and noise measurements.

The spectral response measurements were carried out with a Fourier Transform Infrared Spectrometer or FTIR (Bruker, Equinox IFS55). The transmission of the cryostat window was taken into account. The spectral response measurements are presented in section 5.

Responsivity measurements were realized by means of a CI SR 80 blackbody, with a 33° field of view.

To complete these experiments, noise measurements were performed. Because of the wide frequency range (1 to $10^3$ Hz), a particular care is required to prevent electromechanical parasitic noise. Thus we used a mechanical decoupling system between the ground and our cryostat. Specific doubly shielded wires were also required. The noise signal was sent to a Femto–DLPCA 200 transimpedance amplifier with a tunable gain from $10^2$ to $10^9$. The data was acquired using a spectrum analyser (Ono Sokki – CF5220Z).

## 4. Dark current measurements

Fig. 1 presents the dark current measurements as a function of the operating temperature and for bias voltage between 0.5V and 2.5V. The curve exhibits two regimes: for T>35K, the dark current is strongly dependent on the temperature, which is consistent with the expected thermionic regime; for T<25K, a plateau appears: decreasing the temperature further has no effect on the dark current. The absence of dependence of the dark current on the temperature is the signature of the tunnel regime. A plot of the current as a function of 1000/T allows one to extract the activation energy ($E_a$)) in the range of T>35K. For our QWIP, the value of 81meV was extrapolated for the activation energy under null bias. This experimental value is consistently included between $V_b - E_1 = 86 meV$ (corresponding to the transition from the lower populated state to the continuum) and $V_b - E_{fw} = 76 meV$ (corresponding to the transition from the upper populated state to the continuum).

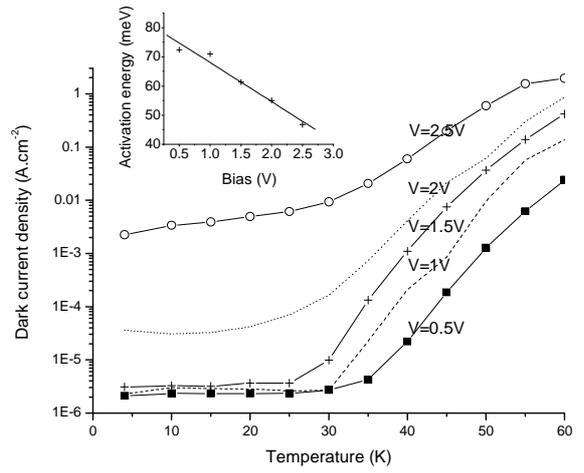

Figure 1. Dark current density as a function of the temperature for bias voltage values ranging from 0.5V to 2.5V by steps of 0.5V. Inset: activation energy as a function of the bias (dots) and linear fit (solid curve).

The same dark current measurements were also plotted as a function of the applied bias for a temperature in the 4K-60K range (see Fig. 2). These I(V) curves exhibit clearly two different regimes: at high temperature (T>25K), a strong dependence on the temperature is observed; whereas under 25K all I(V) curves are superimposed, which is the evidence that the tunnel regime has been reached. It corresponds to the plateaus in the current-temperature curves of Fig. 1. The low temperature I(V) curves can be split into three regimes: first an ohmic regime (V<0.5V) with a linear increase of the current with the bias. For higher bias (0.5V<V<1.5V) the I(V) curves present a plateau behavior, where the dependence with the bias voltage is surprisingly low. Such a low dependence on bias voltage has already been observed on photocurrent I(V) curves [9] (they were attributed to electric field domain formation under illumination). It also has been observed on dark current, see ref [5], [10] and [11], but to the best of our knowledge, such a plateau behaviour has never been deeply investigated. We underline that the coupling in this QWIP (miniband width of a few ten nanoeV) is particularly weak and we do not expect to be on the same regime of transport as much coupled superlattice [10]. It is also interesting to underline that neither hysteresis, nor saw tooth pattern have been observed in those I(V) curves. The plateau current value stays several orders of magnitude above the setup current resolution, which allows us to study this fine effect. For bias voltage higher than 1.5V, a rapid increase with the bias is observed. This rise, which necessarily involves a strong electric field on the first barrier in the vicinity of the contact [12], can be further attributed in the center of the structure to current from ground state to continuum. Indeed for a bias voltage of 1.5V, the potential drop over one period becomes equivalent to the ground-to-excited states transition energy (~80meV). Thus, the resonant transport from the ground state of a well to the excited state of the next well may also occur. This resonance may be broadened by the resonance with other states in the continuum.

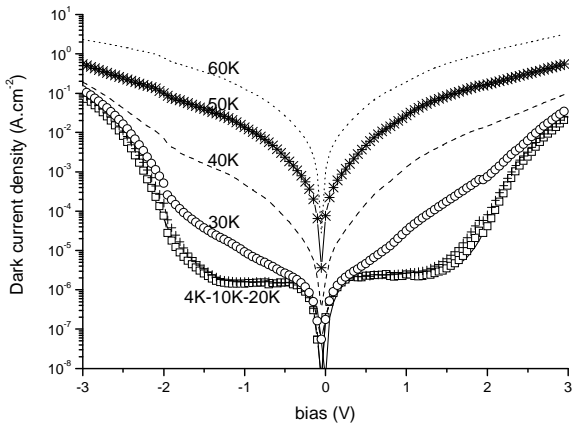

Figure 2. Dark current as a function of the applied bias for different temperatures.

The I(V) curves in Fig. 2 also exhibit an asymmetry with the polarity of the bias. The ratio I+/I- between the currents at +1V and -1V bias voltages is typically of 1.5. This asymmetry was attributed to the doping segregation [13]. The possibility of an asymmetric aluminium profile, due for example to an aluminium overshoot while opening the Al cell, could also be considered. Indeed, the phenomenon is not so simple, since for low temperature the ratio I+/I- is higher than unity at low bias, while it becomes lower than unity at high bias.

## 5. Spectral response, responsivity and noise measurements

Spectral response measurements are reported in Fig. 3. As expected, the peak wavelength is typically 14.5µm. The high wavelength cut-off presents a dependence on the bias voltage, which is attributed to photon and electric field-assisted tunneling, see inset of Fig. 3 [14]. This point will be further discussed in Section 7.

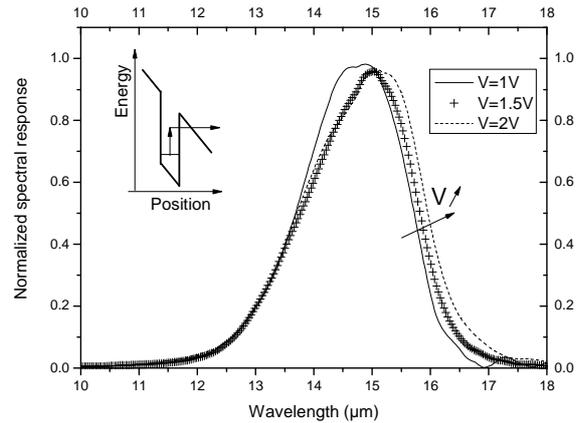

Figure 3. Normalized spectral response as a function of the wavelength for different values of the bias. Inset : explanation of the tunnel effect assisted by photon and electric field.

Responsivity measurements allowed us to determine the external quantum efficiency (product of the internal quantum efficiency by the photoconductive gain): 5% under 1V (responsivity of 0.58A.W$^{-1}$) and 18% under 2V of bias (responsivity of 2.1A.W$^{-1}$) for a temperature of 20K and a wavelength of 14.5µm. Such responsivities values are quite good for QWIP [15]. QWIP optimisation always requires a trade-off between dark current and responsivity. Since 15µm QWIP are operated in very specific conditions (low operating temperature, low incident power) this trade-off necessarily leads to a specific optimisation concerning the doping. For this purpose, a full understanding of electronic transport in these specific structures is necessary, as will be shown later.

Finally, we measured the noise spectral density, under dark condition. We were able to measure a current noise density as low as $3 fA/\sqrt{Hz}$, for T under 20K, in a range of frequency (400-800Hz) where the noise spectral density presents a plateau. To evaluate the detectivity, we only considered the noise due to the dark current since we are considering low flux and thus low photocurrents. We thus obtain a detectivity, under dark condition, as high as 5.2 10$^{12}$ Jones at 20K (3.10$^{10}$ Jones at 50K) and under 2V bias voltage.

# 6. Comparison of dark current measurements with theory

In this section we compare our dark current measurements with present theories. We start by summarizing the main expressions that can be found in the literature, together with their physical background, and then compare their theoretical predictions with our experimental results.

*A. High temperature regime modelling*

Different expressions of the thermionic current ($J_{th}$) can be found in the literature. The expressions (1), (2) and (3) are respectively due to Levine [8], Schneider and Liu [15] and Kane [16].

$$J_{th\ Levine} = e \cdot \frac{m^*}{\pi \hbar^2} \cdot \frac{v_d}{L_w} \cdot eFL_b \cdot e^{-\frac{V_b - eFL_w - E_{fw}}{k_b T}} \quad (1)$$

$$J_{th\ SchneiderLiu} = e \frac{m^*}{\pi \hbar^2} \frac{v_d \tau_c}{\tau_{scatt}} \frac{k_b T}{L_d} e^{-\frac{V_b - \frac{eFL_w}{2} - E_{fw} + E_{ex}}{k_b T}} \quad (2)$$

$$J_{th\ Kane} = 2ev_d \left(\frac{m^* k_b T}{2\pi \hbar^2}\right)^{3/2} e^{-\frac{V_b - \frac{eFL_w}{2} - E_{fw} + E_{ex}}{k_b T}} \quad (3)$$

where e is the elementary charge, m* the effective mass of GaAs, $\hbar$ the reduced Planck constant, $v_d$ the drift velocity, $\tau_c$ the capture time, $\tau_{scatt}$ the scattering time from ground state to continuum, $L_d$ the period width and $E_{ex}$ the exchange energy [17]. The value of $E_{ex}$~-13.6mev is obtained following the approximate expression from ref [18]:

$$E^{ex} \approx -\frac{e^2}{4\pi\varepsilon} k_F (1 - 0.32 \frac{k_F L_w}{\pi}) \quad (4)$$

with $k_F$ the Fermi wave vector and $\varepsilon$ the GaAs permitivity.

These current expressions all rely on an Arrhenius law $I_{Th} \prec e^{-\frac{E_a}{k_b T}}$, with $k_b$ the Boltzmann constant, T the temperature. The term $e^{-\frac{E_a}{k_b T}}$ can either be seen as a population factor near the continuum energy or as a probability for an electron on the ground state to be scattered to the continuum. However, these expressions are quite different since both the expression of the activation energy and the factors before the exponential are different. Considering the activation energy, a bias dependence is introduced via a factor $eFL_w$. Such a dependence reflects the effective barrier lowering due to the electric field, assuming that no Stark shift will affect the ground state energy. The $\frac{eFL_w}{2}$ dependence, preferred in equations (2) and (3) is equivalent to considering that the electrons are globally in the center of the well. The drift velocity also includes a bias dependence: $v_d = \frac{\mu F}{\sqrt{1 + \left(\frac{\mu F}{v_{sat}}\right)^2}}$ with $\mu$ the mobility and $v_{sat}$ the saturation velocity. Both μ and $v_{sat}$ are adjusting parameters. Typical values can be found in the literature, for example in ref. [19]: μ=0.1 m$^2$.s$^{-1}$.V$^{-1}$ and $v_{sat}$=5.10$^4$ m.s$^{-1}$.

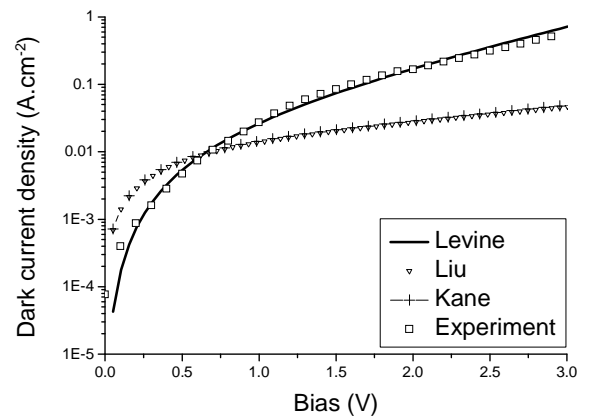

Figure 4. Comparison of the different current expressions as a function of the bias at 50K.

Following [15] the parameter $\frac{\tau_c}{\tau_{scatt}}$ is taken equal to unity.

Fig. 4 shows the comparison of the different current expressions with the experimental current. Liu and Kane expressions give very similar results whereas their prefactor expressions are quite different. For our QWIP the best agreement with the experimental current is obtained by Levine expression. In particular the extra dependence with the electric field in the prefactor leads to a better agreement. Moreover the high number of adjusting parameters in (2) and (3) makes these equations less trivial to use for any structure. Therefore, we will only consider the first expression from now on. The following physical picture can be used to interpret Eq.(1): the dark current is the product of the elementary charge by a three dimensional (3D) carrier density and by a velocity. The 3D carrier density is built from a two dimensional (2D) density of states divided by the well width. Finally, this 3D density of states is multiplied by the typical energy of the problem $eFL_b$ to give a 3D carrier concentration.

B. *Low temperature regime modelling*

The low temperature current is mostly attributed to sequential tunneling. In this regime, electrons tunnel through the barrier. The modelling of this tunneling requires the use of a transmission coefficient often evaluated using the Wentzel-Kramers-Brillouin (WKB) approximation. Akin to the high temperature regime, several expressions of the current can be found in the literature. Equation (5) was derived by Levine [8], (6) by Gomez [20] et al, and (7) by Schneider and Liu [15].

$$J_{ST\ Levine} = \frac{ek_bT}{\hbar L_w^2}.P(E_1+eFL_w)$$

$$\times \ln\left(\frac{1+e^{\frac{E_{fw}-E_1}{k_bT}}}{1+e^{\frac{E_{fw}-eFL_b-E_1}{k_bT}}}\right) \quad (5)$$

$$J_{ST\ Gomez} = \frac{ek_bT}{2\hbar L_w^2}.P(E_1)$$

$$\times \ln\left(\frac{1+e^{\frac{E_{fw}-E_1}{k_bT}}}{1+e^{\frac{E_{fw}-eFL_b-E_1}{k_bT}}}\right) \quad (6)$$

$$J_{ST\ SchneiderLiu} = e\frac{m^*}{\pi\hbar^2}\int_{E1}^{\infty}\tau^{-1}(E)f_{FD}(E)dE$$

$$- e\frac{m^*}{\pi\hbar^2}\int_{E1-eFL_b}^{\infty}\tau^{-1}(E)f_{FD}(E)dE \quad (7)$$

Here P is the WKB probability for an electron to cross the trapezoidal part of the barrier and is given by

$$P(E) = \exp\left(-\frac{4\sqrt{2m_b^*}}{3eF\hbar}\left[(Vb-E)^{3/2} - (Vb-eFL_b-E)^{3/2}\right]\right) \quad (8)$$

Expressions (5) and (6) in fact result from simplifications of equation (7). In this last expression the current is the product of the elementary charge by the 2D density of states, and by a time constant associated to the electron tunnelling through the barrier. This quantity is integrated over all initial states and weighted by the Fermi Dirac ($f_{FD}$) population factor of each state. The current expression is the sum of two terms: the first one represents the current which goes down the structure. The second term is the current which flows back from the lower energy quantum well. The tunneling time constants can be written with a "ping pong" approach [21, 22, 23], where the electron goes from one side of the well to the other in a time equal to two well widths (one well width in Levine et al. [8] expression) divided by the electron velocity. Each time the electron reaches a side of the well, it gets a probability P to tunnel through it. Thus the tunneling time constant is given by $\tau = \frac{2L_w}{v_1}P^{-1}$, with a factor two if we consider that only one side of the well contributes to the current. $v_1$ is then linked to the energy by $E = 1/2.m^*v_1^2$. A key approximation to obtain the expressions (5) and (6) is to consider that the tunneling time constant is independent of the energy, in order to extract the tunnelling rate of the integral. Such a hypothesis is a better approximation for low doped QWIP for which the electronic population has a limited extent in energy. Expressions (5) and (6) mostly differ by the value at which the probability to cross the barrier is evaluated. Levine also increases the field dependence by introducing a factor $eFL_w$. Such an extra dependence is not in favour of the plateau prediction. Then the integration of the Fermi-Dirac factor and the use of log property leads to the supply function $k_bT \ln\left(\dfrac{1+e^{\frac{E_{fw}-E_1}{k_bT}}}{1+e^{\frac{E_{fw}-eFL_b-E_1}{k_bT}}}\right)$. In this expression, a factor $eFL_d$ could be preferred to $eFL_b$ for the population of the lower well, but the difference will be quite low since $L_b \gg L_w$.

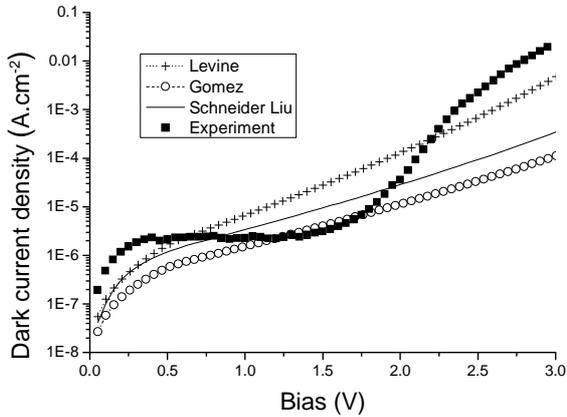

Figure 5. Comparison of the different expressions for the tunneling current as a function of the bias, at T=10K.

Fig. 5. presents the current associated with the different tunnelling current expressions. Expressions (6) and (7) give very similar shape, and they only differ by a factor two (due to the fact that (7) includes the energy dependence of the tunnelling probability). As expected (5) exhibits a higher bias dependence due to the factor $eFL_w$. It is difficult to determine which expression gives the better agreement with the experiment, since none of them reproduces correctly the plateau behavior. In the following we have chosen to use the expression (7) rather than the simplified ones, particularly because our doping is not so low.

Finally the global theoretical current can then be written as the sum of the thermionic ($J_{th}$ – expression (1) ) and the sequential tunneling ($J_{st}$ – expression (7) ) contributions. A comparison of the experimental and theoretical I(V) curves is shown in Fig. 6(a) (high temperature) and Fig. 6(b) (low temperature). The high temperature modelling presents quite a good agreement with the experimental dark current, particularly at low bias voltage (V<2V). The disagreement at high bias voltage (V>2V) is due to the change of transport mechanism, not taken into account in our current expression. At low temperature, because of the superposition of experimental I(V) curves, we only have plotted the case for T=10K, see Fig. 6(b). The agreement between experimental and theoretical current is low. *Indeed, neither the order of magnitude nor the shape are correctly reproduced.*

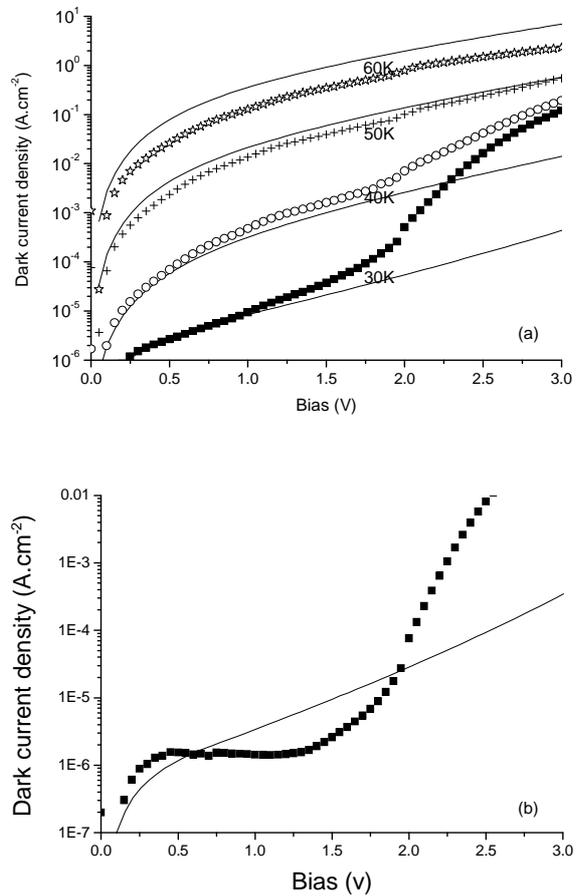

Figure 6. (a) Theoretical dark current density (solid line) and experimental (scatter line) as a function of the applied bias for temperatures from 30K to 60K by step of 10K. (b) Theoretical (solid line) and experimental (scatter line) dark current density as a function of the applied bias at 10K.

## 7. Is the plateau regime a consequence of electric field domain formation?

*The plateau regime in the I(V) curves clearly corresponds to a situation where a change in the bias voltage has no effect on the current flowing through the structure.* A possible explanation of this phenomenon could be the following: since the current is controlled by the electric field reigning at the contact [24], this surface electric field could stay unchanged when the bias voltage increases, while the potential drop is localized in one or a few periods of the structure. This would result in high electric field domain formation either in the vicinity of the contact [25] or in the bulk of the structure [9]. However, when electric field domains are formed, the I(V) curve presents saw tooth patterns or equivalently peaks in the conductance curve [10, 26]. This is not the case in our sample. Moreover the spectral response measurements reveal that the high wavelength cut-off increases with the bias voltage, which gives strong indication that the potential drop is not concentrated only on a few periods, but distributed over the whole structure [21]. We underline the fact that this distribution is not necessarily homogeneous.

## 8. Conclusion

The tunneling regime undoubtedly gives access to the ultimate performances of QWIP detectors in terms of dark current. Because the dark current is a crucial issue for VLWIR low background applications, we have studied its behaviour with great attention in 15 µm-peaking devices. We have evidenced a striking feature in I(V) curves at low temperature, i.e. a plateau regime. Let us note that this plateau behaviour is not sample dependent and is observed in samples presenting different technological parameters.

Simple theoretical expressions allowed us to reproduce I(V) curves at high temperature with a good agreement. However, we have shown that these simple expressions are not able to describe the I(V) curves at low temperature. In particular, the plateau structure in the I(V) curves cannot be explained with the WKB approximation. Spectral response measurements allowed us to show that this plateau could not be attributed to high electric field domain formation either.

In other words, the available theoretical tools don't allow us to understand the physical phenomena involved in dark current in the tunnel regime. We believe that this theoretical problem could be solved thanks to a microscopic model using a scattering approach. Thus, all the relevant interactions could be taken into account. Such a microscopic modelling of the dark current is absolutely necessary to identify the major contribution to dark current, and to finally improve the design of QWIP structure.

## References

[1] S.D. Gunapala, S. V. Bandara, J. K. Liu, J. M. Mumolo, S. B. Rafol, D. Salazar, J. Woolaway, P. D. LeVan and M. Z. Tidrow, Infrared Phys. Techn. **50** (2007) 217-226.
[2] M. Jhabvala, K. K. Choi, C. Monroy and A. La, Infrared Phys. Techn. **50** (2007) 234-239.
[3] G. Sarusi, S.D. Gunapala, J.S. Park, B. F. Levine, J. Appl. Phys. **76** (1994), 6001-6008.
[4] A. Nedelcu, N. Brière de l'Isle, J-P. Truffer, E. Belhaire, E. Costard, P. Bois, P. Merken, O. Saint-Pé, SPIE 7106-57 (to be published).
[5] A. Singh and D.A. Cardimona, Proc. SPIE vol. 2999.
[6] E. Costard, A. Nedelcu, M. Achouche, J. L. Reverchon, J. P. Truffer, O. Huet, L. Dua, J. A. Robo, X. Marcadet, N. Brière de l'Isle, H. Facoetti, and P. Bois, Proc. SPIE, Vol. 6744 (2007), 674411.
[7] E. Costard and Ph. Bois, Infrared Phys. Techn **50** (2007), 260-269.
[8] B.F. Levine, J. Appl. Phys. **74** (1993), R1.
[9] H.Schneider, C. Shönbein, R. Rehm, M. Walther, P. Koidl, App. Phys. Lett. **88** (2006), 051114.
[10] K.-K. Choi, B. F. Levine, C. G. Bethea, J. Walker, and R. J. Malik, App. Phys. Lett. **50** (2005), 1814.
[11] H. Martijn, A Gromov, S. Smuck, H. Malm, C. Asplund, J. Borglind, S. Becanovic, J. Alverbro, U. Halldin, B. Hirschauer, Infrared Phys. Techn. **47** (2005), 106.
[12] L. Gendron, V. Berger, B. Vinter, E. Costard, M. Carras, A. Nedelcu and P. Bois, Semiconductor Science and Technology **19** (2004), 219.
[13] H. C. Liu, Z. R. Wasilewski, M. Buchanan and Hanyou Chu, Appl. Phys. Lett. **63** (1993), 761.
[14] J. Le Rouzo, I. Ribet, R. Haidar, N. Guérineau, M. Tauvy, E. Rosencher and S.L. Chuang, Appl. Phys. Lett. **88** (2006), 091117.

[15] H. Schneider and H.C. Liu, in *Quantum well infrared photodetectors – Physics and applications* (Springer, Heidelberg, 2006).

[16] M.J. Kane, S. Millidge, M.T.Enemy, D. Lee, D.R.P. Guy and C.R.Whitehouse, in *Intersubband transitions in quantum wells*, (Plenum, New York, USA, 1992).

[17] G. E. W. Bauer and T. Ando, Phys. Rev. B **34** (1986), 1300.

[18] K.M.S.V. Bandara, D.D. Coon, O. Byungsung, Y.F. Lin, M.H. Francombe, App. Phys. Lett. **53** (1989), 1931.

[19] L. Thibaudeau, P. Bois and J.Y. Duboz, J. Appl. Phys. **79** (1996), 446.

[20] A. Gomez, V. Berger, N. Péré-Laperne and L.D. Vaulchier, App. Phys. Lett. **92** (2008), 202110.

[21] E. Martinet, E. Rosencher, F. Chevoir, J. Nagle and P. Bois, Phys. Rev. **44** (1991), 3157.

[22] J.R. Oppenheimer, Phys. Rev. **31** (1928), 66.

[23] Eyvind H. Wichmann., in Berkeley physics courses: Quantum physics, vol. 4 (McGraw-Hill, New york, 1998).

[24] E. Rosencher, F. Luc, P. Bois and S. Delaitre, Appl. Phys. Lett. **61** (1992), 468.

[25] M. Ershov, V.Ryzhii, C. Hamaguchi, Appl. Phys. Lett. **67** (1995), 3147.

[26] L. Esaki and L. L. Chang, Phys. Phys. Lett. **33** (1974),495.

**Figure captions**

Fig. 1. Dark current density as a function of the temperature for bias voltage values ranging from 0.5V to 2.5V by steps of 0.5V. Inset: activation energy as a function of the bias (dots) and linear fit (solid curve).

Fig. 2. Dark current as a function of the applied bias for different temperatures.

Fig. 3. Normalized spectral response as a function of the wavelength for different values of the bias. Inset : explanation of the tunnel effect assisted by photon and electric field.

Fig. 4. Comparison of the different current expressions as a function of the bias at 50K. Following [15] the parameter $\dfrac{\tau_c}{\tau_{scatt}}$ is taken equal to unity.

Fig. 5. Comparison of the different expressions for the tunneling current as a function of the bias, at T=10K.

Fig. 6. (a) Theoretical dark current density (solid line) and experimental (scatter line) as a function of the applied bias for temperatures from 30K to 60K by step of 10K. (b) Theoretical (solid line) and experimental (scatter line) dark current density as a function of the applied bias at 10K.